\documentclass[12pt]{article}
\usepackage[dvips]{graphicx}
\usepackage{latexsym}
\usepackage{amssymb}
\newcounter{mnotecount}[section]

\renewcommand{\themnotecount}{\thesection.\arabic{mnotecount}}

\newcommand{\mnote}[1]
{\protect{\stepcounter{mnotecount}}$^{\mbox{\footnotesize
$
\bullet$\themnotecount}}$ \marginpar{
\raggedright\tiny\em
$\!\!\!\!\!\!\,\bullet$\themnotecount: #1} }

\newcommand{\R}{\mathbb{R}}

\def\p{\partial}

\newcommand{\bea}{\begin{eqnarray}}
\newcommand{\eea}{\end{eqnarray}}

\def\be{\begin{equation}}
\def\ee{\end{equation}}

\def\const{\mbox{const}}

\def\p{\partial}

\renewcommand{\themnotecount}{\thesection.\arabic{mnotecount}}
\begin{document}

\title{\vskip -70pt
\begin{flushright}
{\normalsize DAMTP-2012-1} \\
\end{flushright}
\vskip 10pt
{\bf Abelian vortices from Sinh--Gordon and Tzitzeica equations
\vskip 10pt}}
\author{Maciej Dunajski\thanks{Email: M.Dunajski@damtp.cam.ac.uk}
\\
Department of Applied Mathematics and Theoretical Physics,\\
University of Cambridge,\\
Wilberforce Road, Cambridge CB3 0WA, UK. }
\date{}
\maketitle
\begin{abstract}
It is shown that both the sinh--Gordon equation and the elliptic
Tzitzeica equation can be interpreted as the Taubes equation for
Abelian vortices on a CMC surface embedded in $\R^{2, 1}$, or on a surface
conformally related to a hyperbolic affine sphere in $\R^3$. 
In both cases the Higgs field and the $U(1)$ vortex
connection are constructed directly from the Riemannian data of the surface
corresponding to the sinh--Gordon or the Tzitzeica equation. Radially symmetric
solutions lead to vortices with a topological charge equal to one, 
and the connection formulae for the resulting third Painlev\'e transcendents
are used to compute explicit values for the strength of the vortices.
\end{abstract}
\section{Introduction}
The Abelian Higgs model is a relativistic field 
theory  of a complex scalar field $\phi$ coupled to an Abelian gauge field $a$ on a 2+1 dimensional space--time $\Sigma\times\R$.
Here $\Sigma$ is a two--dimensional surface with a Riemannian metric
$g$ and the space--time metric is $g-dt^2$. This model admits topological solitons called vortices \cite{MS} relevant in the theory of thin superconductors \cite{abrikosov}.

At the `critical coupling'  the vortices arise 
as solutions to the first order Bogomolny equations. These Bogomolny
equations are in general not integrable and the form
of static finite energy solutions is not known. This sets the vortices
apart from other topological solitons (lumps, monopoles, Yang--Mills 
instantons), where the Bogomolny equations are integrable and large
families of explicit solutions exist \cite{Dunajski_book}. The one exception is when the the
surface $(\Sigma, g)$ is a hyperbolic space with a metric of constant
curvature $-1/2$. In this case the Bogomolny equations reduce to the Liouville equation whose solutions are known \cite{Wi, Strachan, MR}.

 The aim of this note is to show that there are two more 
integrable cases. In both cases the Higgs field and the connection are  constructed directly from the Riemannian data of the surface $(\Sigma, g)$. Thus, unlike the hyperbolic vortex, the resulting solutions are 
isolated points in the moduli space: changing the parameters of a vortex would correspond 
to changing the background metric on $\Sigma$. The two integrable 
cases  
correspond to $(\Sigma, g)$ being a surface of constant mean curvature
in $\R^{2,1}$ and a surface conformal to  a 
hyperbolic affine sphere in $\R^{3}$ respectively.
In both cases the norm of the Higgs field
is given by a power of the conformal factor in the metric $g$, and the $U(1)$ connection
is identified with a Levi--Civita connection one--form of $g$. 
The vortex number is given by the Euler characteristics of $T\Sigma$.
In the first case the Bogomolny equations
reduce to an elliptic sinh--Gordon equation, and in the second case
they reduce to an elliptic Tzitzeica equation. Both
sinh--Gordon and Tzitzeica equation arise as symmetry reductions
of anti--self--dual Yang--Mills equations in $\R^4$ and thus are integrable \cite{D02,Dunajski_book}. 
The radial solutions of sinh--Gordon and Tzitzeica equations 
are characterised by 3rd Painlev\'e
transcendents with parameters $(0, 0, 1, -1)$ and $(1, 0, 0, -1)$ respectively. 
Using the known connection formulae \cite{MTW77,Kit0,Kit} for 
these transcendents relating asymptotic of the solutions at $0$ and $\infty$ we are able to find analytic expressions for strengths of the corresponding one--vortex solutions.
\section{Taubes equation}
Let $L$ be a Hermitian complex line bundle over a Riemannian surface
$(\Sigma, g)$ and let $\omega$ be a K\"ahler form corresponding to
some choice of a complex structure on $\Sigma$.
We define vortices as finite energy solutions 
of the Bogomolny equations
\[
\bar{D}\phi=0, \quad F=\frac{1}{2}\omega(1-|\phi|^2),
\]
where the Higgs field $\phi$ is a global $C^{\infty}$ section of $L$,
$\bar{D}$ is the anti--holomorphic part of the covariant derivative
of a $U(1)$ connection $a$ on $L$ compatible with the Hermitian structure of 
$L$, and finally $F$ is the curvature of $a$. The energy functional
is given by
\[
E[a, \phi] =\frac{1}{2}\int_{\Sigma} \Big(|D\phi|^2+|F|^2+\frac{1}{4}(1-|\phi|^2)^2\Big)
\mbox{vol}_\Sigma.
\]
Let $z=x+iy$ be a local
holomorphic coordinate on $\Sigma$ such that the metric takes
the form
\[
g=\Omega \;dz d\bar{z},\quad\mbox{where}\quad \Omega=\Omega(z, \bar{z}).
\]
Choosing a trivialisation of $L$ such that $a=a_zdz+\bar{a}_zd\bar{z}$,
the curvature is given by the magnetic field $B=\p_x a_y-\p_y a_x$
and the Bogomolny equations become
\be
\label{first_bog}
\frac{\p \phi}{\p \bar{z}}=i a_{\bar{z}}\phi
\ee
and 
\be
\label{second_bog}
B=\frac{\Omega}{2}(1-|\phi|^2).
\ee
We solve the first equation for $a$ and set 
$\phi=\exp{(h/2+i\chi)}$, where $h, \chi:\Sigma\rightarrow \R$.
The second equation  then yields the  Taubes equation \cite{Taubes_r}
\be
\label{Taubes}
\Delta_0 h+\Omega-\Omega e^h=0, \quad\mbox{where}\quad  
\Delta_0=4\p_z\p_{\bar{z}}.
\ee
This is valid outside small discs enclosing the 
logarithmic singularities of $h$. 
The vortices are centred at points in $\Sigma$
where the Higgs field vanishes or equivalently where 
$h\rightarrow -\infty$.
The vortex number is defined to be
\[
\frac{1}{2\pi}\int_{\Sigma} B\;\mbox{vol}_{\Sigma},
\]
and an $N$--vortex solution centred at $z=z_0$ has an expansion of the form
\be
\label{assymp_t}
h\sim 2N\log|z-z_0|+\mbox{const}+\frac{1}{2}\bar{b}(z-z_0)
+\frac{1}{2}{b}(\bar{z}-\bar{z}_0)+\cdots
\ee
as $|z|\rightarrow 0$, where the coefficients
expansion depend on the position of the vortex.
The  moduli space of solutions to  (\ref{first_bog}) and (\ref{second_bog}) with vortex number $N$ is a manifold of real dimension $2N$.

\section{Vortices from sinh--Gordon equation}
 Taking $\Omega=\exp{(-h/2)}$ in the Taubes equation gives
the elliptic sinh--Gordon equation
\be
\label{sinh}
\Delta_0\Big(\frac{h}{2}\Big)=\sinh\Big(\frac{h}{2}\Big).
\ee
This gives an interpretation of the metric $g$  as an {
isolated vortex}. The magnetic field and the Higgs field
have an intrinsic geometric interpretation as the (Hodge dual of) the  Riemann curvature two--form and the (inverse of) conformal 
factor with a complex phase. The Higgs field vanishes at
the position of the vortex and at this point the conformal factor
becomes infinite. The norm of the Higgs field $|\phi|^2$ tends to its 
asymptotic value $1$ as $x^2+y^2\rightarrow \infty$.

  The first Bogomolny equation
(\ref{first_bog}) asserts that the $U(1)$ connection $a$ is
 equal to the Levi--Civita connection one--form
of $g$. This is true in some trivialisation
of the underlying line bundle $L$ identified with the tangent bundle
$T\Sigma$. The holonomy group
$SO(2)$ of the Levi--Civita connection is then identified with the gauge group $U(1)$.
The Chern number of $L$ becomes the Euler characteristic of $T\Sigma$.

 The second Bogomolny equation (\ref{second_bog})
is the sinh-Gordon equation (\ref{sinh}). It imposes conditions
on the metric $g$ which are equivalent to
the statement that the background surface 
$(\Sigma, g)$ is a space--like immersion with constant mean curvature
(CMC)
in the flat Lorentzian three--space $\R^{2, 1}$. In this context
equation (\ref{sinh}) arises from the Gauss--Codazzi 
equations \cite{brander,Dorf,Yamada}.

To see the explicit relations between the Riemannian data of $g$ and
the vortex, set $h=-2u$ so that
\begin{eqnarray}
\label{CMC_1}
&&\Delta_0(u)=\sinh(u), \quad g=e^udzd\bar{z}, \quad\mbox{and}\nonumber\\
&& B=\Delta_0(u), \quad \phi=e^{-u+i\chi}, \quad
a=i(\bar{\partial}-\partial) u-d\chi,
\end{eqnarray}
where $\partial =dz\otimes\p/\p z$.
Choose a spin--frame ${\bf e}^1=e^{u/2}dx, {\bf e}^2=e^{u/2}dy$
such that $g=\delta_{ij}{\bf e}^i\otimes {\bf e}^j$.
The connection
and curvature forms of $g$ can be read off from the Cartan 
structure equations and are given by
\begin{eqnarray*}
{\Gamma^{1}}_2&=&\frac{1}{2}(u_ydx-u_xdy)=-\frac{1}{2}a-\frac{1}{2}d\chi,\\ 
{R^1}_2&=&-\frac{1}{2}(u_{xx}+u_{yy})dx\wedge dy=
-\frac{1}{2}B dx\wedge dy. 
\end{eqnarray*}
  A simple solution to (\ref{sinh}) with $h=-2u$ is
\be
\label{soliton}
u=4\tanh^{-1}\Big(\exp{\Big(\frac{(z-z_0)e^{-i\alpha}}{2}+ 
\frac{(\bar{z}-\bar{z}_0)e^{i\alpha}}{2}\Big)}\Big),
\ee
where $\alpha$ is a real parameter. This is an analytic continuation
of the one-kink soliton of the better known sine--Gordon equation.
Setting $\alpha=0, z_0=0$ gives the metric
on the surface of revolution
\[
g=\tanh{(x/2)}^{-2}(dx^2+dy^2).
\]
Expanding this around $x=0$ gives
the half plane metric
\[
\frac{1}{4}\frac{dx^2+dy^2}{x^2}.
\]
The expansion of $h$ around 0 is $4\log|x|+\cdots$ which 
however can not be interpreted as a two--vortex solution
as $B$ doesn't vanish in the $y$ direction and the integral is infinite.
\vskip5pt
To reinterpret the CMC surface (\ref{CMC_1}) as a vortex we need to construct a  solution of $\Delta_0{u}=\sinh{(u)}$ such that
\[
u\rightarrow 0 \quad\mbox{as}\quad r\rightarrow \infty
\]
and $h=-2u$ behaves like (\ref{assymp_t}) near $r=0$.
We shall look for radial solutions of the form $u=u(r)$. The sinh--Gordon 
equation reduces to an ODE
\be
\label{PIII}
u_{rr}+\frac{1}{r}u_r=\sinh{u}.
\ee
This is equivalent to the Painlev\'e III ODE with special values of parameters.
For large $r$ we approximate $\sinh{u}=u$, and obtain the modified Bessel  equation of order zero, so
\be
\label{as1}
u\sim 4\lambda K_0(r)\qquad\mbox{as}\quad r\rightarrow \infty,
\ee
for some $\lambda$ which needs to be determined by the initial condition at $r=0$.
For small $r$ we instead get the Liouville equation
\be
\label{rsg}
u_{rr}+\frac{1}{r}u_r=\frac{1}{2}e^{u}
\ee
which can be solved exactly. Thus
\be
\label{as2}
u\sim -2\sigma\log(r)+\const+O(r) \qquad\mbox{as}\quad r\rightarrow 0
\ee
for some constant $\sigma$ (the overall multiples 2 and 4 in these
asymptotic have been chosen for convenience).
It is known \cite{Kaup} that if $0<\lambda<\pi^{-1}$ there exists a solution to the radial sinh--Gordon equation whose only singularity is at $0$.  To find how $\lambda$ depends on $\sigma$ we shall
refer to the results about the Painlev\'e III asymptotics.
Setting
$u=-2\ln{w}$ and $r=2\rho$ in (\ref{rsg}) yields
\[
w_{\rho\rho}=\frac{(w_\rho)^2}{w}-\frac{w_\rho}{\rho}+w^3-\frac{1}{w},
\]
which is the Painlev\'e III equation
with parameters  $(0, 0,  1, -1)$. 

The connection formulae relating the asymptotic solution
to PIII at $r=0$ and $r=\infty$ have been derived in \cite{MTW77}. 
Using the asymptotic formula for the modified Bessel function
\be
\label{bessel_as}
K_0(r)=\sqrt{\frac{\pi}{2}}\frac{1}{\sqrt{r}}e^{-r}
\ee
valid for large $r$ and applying the results\footnote{We need formula (1.10) from this reference with $u(r)=-2\log{(w(r/2))}$, 
$\nu=0$ and expressions (1.11) and (1.12) from {\it Theorem 3}.}
of \cite{MTW77}  we find
that $h=-2u$ is of the form
\begin{eqnarray}
\label{assymp_h}
h(r)&\sim&4\sigma \ln{r}+4\ln{\beta}+\frac{1}{\beta^2}r\quad \mbox{for}\quad  r\rightarrow 0\\
&\sim& -8\lambda K_0(r)\quad \mbox{for}\quad r\rightarrow \infty\nonumber
\end{eqnarray}
with the connection formulae
\[
\sigma(\lambda)=\frac{2}{\pi}\arcsin{(\pi\lambda)},
\quad \beta(\lambda)=2^{-3\sigma}\frac{\Gamma((1-\sigma)/2)}{\Gamma((1+\sigma)/2)},
\]
where $\Gamma$ is the gamma function. This is valid for $0\leq \lambda
\leq \pi^{-1}$. The asymptotic expansion (\ref{assymp_t}) implies
that the $N$--vortex solution has $\sigma=N/2$. Thus the asymptotic
connection formulae are valid only if $N=1$, an
there exists a one--vortex solution with $\sigma=1/2$ and
\[
\lambda=\frac{\sqrt{2}}{2\pi}, \quad  \beta=2^{-3/2}\frac{\Gamma(1/4)}{\Gamma(3/4)}.
\]
In the particle interpretation of Speight \cite{speight}, vortices, when viewed from a large distance, behave as  point particles with the strength given by a coefficient of the Bessel function in the asymptotic expansion (\ref{assymp_h}) of $h$.
Thus the strength of the sinh--Gordon vortex is 
\[
\frac{4\sqrt{2}}{\pi}\sim 1.80.
\]
For comparison, the strength of a plane vortex can not be calculated
analytically. The approximate value $1.68$ has been found numerically \cite{MS}.

Near $r=0$ the metric $g$ in (\ref{CMC_1}) with $z=r\exp{i\theta}$
takes the form
\[
g=\frac{4}{\beta^2}(dR^2+ \frac{1}{4}R^2d\theta^2), \quad\mbox{where}\quad
R=r^{-1/2}.
\] 
Thus close to $r=0$ we obtain a  flat metric with conical singularity at the origin
and  
the 1--vortex deficit angle is $\pi$.
The corresponding
CMC surface is an analogue the Smyth CMC surface \cite{smyth}.

 Calculating the vortex number for the solution 
with asymptotic (\ref{as1}) and (\ref{as2}) yields
\begin{eqnarray*}
\frac{1}{2\pi}\int_{\R^2}B\;dxdy&=& 
\frac{1}{2\pi}\lim_{r_2\rightarrow \infty}\lim_{r_1\rightarrow 0}
\int_{0}^{2\pi}\int_{r_1}^{r_2} \frac{1}{r}(ru_r) \;rdrd\theta\\
&=&4\lambda\;\lim_{r_2\rightarrow \infty}(-r_2K_1(r_2))-
\lim_{r_1\rightarrow 0}(-2\sigma+O(r_1))\\
&=&2\sigma
\end{eqnarray*}
as the modified Bessel function $K_1(r)$ decays exponentially at
$\infty$. Thus the vortex number is $1$ if $\sigma=1/2$ as expected.
\subsection{Vortices from Tzitzeica equation}
There is another possible choice of the conformal factor
in (\ref{Taubes}) which leads to an integrable equation. Setting
$h=3u$ and choosing $\Omega=\exp{(-2u)}$ leads 
to the elliptic Tzitzeica equation
\[
\Delta_0 u+\frac{1}{3}(e^{-2u}-e^u)=0.
\]
There are several versions of the Tzitzeica equation which depend on the 
relative signs between the exponential terms \cite{DP08}. 
The one above
corresponds to hyperbolic affine sphere in $\R^3$, \cite{Wang}.
with the Blaschke metric $g_B=e^{3u}g$.

Assuming that $u=u(r)$ and setting (similar reductions have been 
accomplished in \cite{Kit0, DP08})
\[
u(r)=\ln{(w(r))}-\frac{1}{2}\ln{r}+\frac{1}{4}\ln{\Big(\frac{27}{4}\Big)},
\quad r= \frac{3\sqrt{3}}{2}\rho^{2/3}
\]
yields 
\[
w_{\rho\rho}=\frac{(w_\rho)^2}{w}-\frac{w_\rho}{\rho}+\frac{w^2}{\rho}-\frac{1}{w}
\]
which is also Painleve III, this time with parameters
$(1, 0, 0, -1)$ (it may be interesting to note that the Painleve ODEs
also arise in the theory of Chern--Simons--Higgs vortices \cite{schif}). 
The asymptotic connection formulae for this
equation have been obtained in \cite{Kit0}. There exists a 
one--parameter family of solutions singular only at the origin.
Adapting the results of \cite{Kit0} to our case\footnote{We need formulae from
page 2081 with $\varepsilon=1, \tau=r^2/12$. Formula
$(18)$ in Kitaev's paper is used with
$g_1=g_2=0, g_3=1$ and $s-1=2\cos{p}$  with
$0<p<\pi$ so that $\mu=3p/2\pi$.} we find
\begin{eqnarray}
\label{assymp_tzi}
h(r)=3u&\sim& \Big(\frac{9p}{\pi}-6\Big)\log{r}+\beta
\quad \mbox{for}\quad  r\rightarrow 0\\
&\sim& \frac{6\sqrt{3}}{\pi}\Big(\cos{p}+\frac{1}{2}\Big)  K_0(r)\quad \mbox{for}\quad r\rightarrow \infty\nonumber,
\end{eqnarray}
where $0<p<\pi$ parametrises the solutions,
\[
\beta=
3\ln{\Big(3^{-3p/\pi}\frac{9p^2}{2\pi^2}\frac{\Gamma(1-\frac{p}{2\pi})
\Gamma(1-\frac{p}{\pi})}{\Gamma(1+\frac{p}{2\pi})
\Gamma(1+\frac{p}{\pi})}\Big)}-\Big(\frac{9p}{2\pi}-3\Big)\ln{12},
\]
and we have used the 
asymptotic formula for the Bessel function (\ref{bessel_as}).
Comparing this with the $N$--vortex asymptotics
(\ref{assymp_t}) we find that the range of $p$ forces
$N=1$ in which case $p=8\pi/9$ and the strength of the Tzitzeica one--vortex is
\[
|\frac{3\sqrt{3}}{\pi}(2\cos{(8\pi/9)}+1)|\sim 1.45.
\]
The resulting metric near $r=0$ is given by
\[
g=e^{-2u}\;(dr^2+r^2d\theta^2)\sim \mbox{const}\; (dR^2+\frac{1}{9}R^2d\theta^2), \quad\mbox{where}\quad R=r^{1/3}.
\]
Thus near the origin the metric is flat, and has a conical singularity with  the deficit angle $4\pi/3$.
\section{Further remarks}
Static vortices are solitons in two dimensions relevant in the theory of thin superconductors \cite{abrikosov}. The Higgs field $\phi$ describes the density and phase of the
superconducting paired electrons, coupled to an electromagnetic gauge potential $a$. The phase symmetry is spontaneously broken, and consequently electromagnetic field possesses a length scale. In the corresponding relativistic theory this results in the existence of massive photons. The full Ginsburg-Landau
energy functional 
\[
E[a, \phi] =\frac{1}{2}\int_{\Sigma} \Big(|D\phi|^2+|F|^2+\frac{c}{4}(1-|\phi|^2)^2\Big)
\mbox{vol}_\Sigma
\]
leads to second order field equations.
The model analysed in this paper corresponds to the critical value of the coupling constant $c=1$. The case of positive (respectively negative)  $c$
corresponds to Type I (respectively Type II) superconductors. It is known \cite{MS} that certain alloys have $c$ arbitrary close to $1$. The critical coupling is
relevant from the point of view of vortex dynamics, as near the critical coupling the dynamics of slowly moving vortices  can be approximated by a geodesic
motion on the moduli space of static vortices, where the moduli space metric arises from the kinetic term in the vortex Lagrangian.

 The solutions constructed in (\ref{assymp_h} ) and (\ref{assymp_tzi}) correspond to vortices on curved backgrounds $(\Sigma, g)$. The study of such vortices is also physically interesting. For example the case where $\Sigma$ is a flat torus corresponds to periodic Abrikosov lattice of vortices in Type II superconductors \cite{abrikosov}. The general curved backgrounds can be relevant to curved
thin super--conducting materials in three--space.

The case of $\Sigma=\mathbb{H}^2$ with a hyperbolic metric
gives rise to rotationally symmetric Yang--Mills instantons \cite{Wi}. Both the Tzitzeica and Sine-Gordon vortices constructed in this paper can be considered as deformations (albeit isolated - in the sense made clear in the paper) of this example. The corresponding
self--dual Yang Mills equations are satisfied
on a Kahler four--manifold $\Sigma\times S^2$. The scalar--curvature
of this four manifold is non--zero, so this anzatz goes beyond
the analysis of integrable self--dual backgrounds.
\vskip5pt

We have shown that a radially symmetric CMC surface $(\Sigma, g)$ arising
from the elliptic Sinh--Gordon equation (\ref{CMC_1}) can be 
regarded as a 1--vortex
in the Abelian Higgs model. Another 1--vortex corresponds to a radial
solution of an elliptic Tzitzeica equation, and thus to a hyperbolic
affine sphere.

 The vortex number arises  as a topological invariant of the tangent bundle
$T\Sigma$. This is not unlike the recent model proposed in \cite{AMS11},
where the baryon number has been identified with the signature of the underlying 
four--manifold. In \cite{AMS11} the metric connection presumably 
(as the details are not given) gives rise to a gauge potential
whose holonomy is the Skyrme field describing a particle. In our approach
the metric connection on $\Sigma$ leads directly to a vortex. There
may exist other ways  - see \cite{fox} for one possibility - of identifying
the vortex Hermitian line bundle with a tensor power of $T\Sigma$.

 The sinh--Gordon vortex, as well as the Tzitzeica vortex are isolated solutions
as modifying the parameters of vortices would change the underlying 
Riemannian structures $(\Sigma, g)$. The existence of 
moduli spaces is an interesting
problem which we have not been able to resolve.  To understand the nature
of the difficulties let us focus on  the 
sinh--Gordon vortex (\ref{assymp_h}).  
As we have seen, there exist two `obvious'  solutions to the Taubes equation
on the  CMC surface $\Sigma$ with the metric (\ref{CMC_1}): $h=0$ and $h=-2u$. 
Take $u$ to be 
the radially symmetric solution of the sinh--Gordon equation which leads
to a 1--vortex, and consider the perturbed vortex 
$
h=-2u+\phi,
$
where $\phi$ is small. The linearised Taubes equation gives
\[
\triangle_0\phi=e^{-u}\phi
\]
and we stress that this is not the same as the linearised sinh--Gordon 
equation. We require $\phi\rightarrow \mbox{const}$ as $r\rightarrow 0$ and
$\phi\rightarrow 0$ as $r \rightarrow \infty$ to preserve the vortex number.
Looking at the separable solutions gives a radial linear ODE. 
Near $r=0$ this ODE 
admits one regular power series solution. 
For large $r$ the ODE reduces to a modified Bessel equation, 
which admits one solution vanishing at infinity. 
The moduli could  be constructed if we managed to 
relate the overall multiplicative constants in these asymptotic solutions, 
and establish the regularity
for intermediate values of $r$. The metric on the underlying  CMC surface is 
approximately flat for large and
small $r$, but admits a conical singularity at $r=0$ which leads
to a deficit angle depending on a strength of the vortex. Thus it is not clear
whether the results of Taubes \cite{Taubes_r} imply the existence of the moduli 
space.
\subsubsection*{Acknowledgements}
I am grateful  to Josef Dorfmeister, Alexander Kitaev,
Nick Manton and Adam Szereszewski 
for very useful discussions and correspondence. 

\end{document}